\documentclass{article}

\usepackage{spconf}
\usepackage{amsmath}
\usepackage{graphicx}

\usepackage{amsfonts}
\usepackage[english]{babel}
\usepackage{booktabs}
\usepackage{csquotes}
\usepackage[T1]{fontenc}
\usepackage{mathtools}
\usepackage{microtype}
\usepackage{siunitx}
\usepackage{tabularx}
\usepackage{url}

\graphicspath{{figures/}}

\newcommand{\xh}{\hat{x}}
\newcommand{\yh}{\hat{y}}
\newcommand{\Xcal}{\mathcal{X}}
\newcommand{\Ycal}{\mathcal{Y}}
\newcommand{\Xsf}{\mathsf{X}}
\newcommand{\Ysf}{\mathsf{Y}}
\newcommand{\Dc}{\mathsf{D}_C}
\newcommand{\Dn}{\mathsf{D}_N}
\newcommand{\Dr}{\mathsf{D}_R}
\newcommand{\Lq}{\mathcal{L}_q}

\newcommand{\score}[2]{\(#1 \scriptstyle{\pm #2}\)}
\newcommand{\scoreb}[2]{\(\mathbf{#1 \scriptstyle{\pm #2}}\)}
\newcommand{\supstar}[1]{#1\textsuperscript{*}}

\DeclareMathOperator*{\argmax}{arg\,max}
\DeclarePairedDelimiter{\norm}{\lVert}{\rVert}

\title{Learning with Out-of-Distribution Data for Audio Classification}

\name{Turab Iqbal, Yin Cao, Qiuqiang Kong, Mark D. Plumbley, Wenwu Wang}
\address{Centre for Vision, Speech and Signal Processing (CVSSP), University of Surrey, UK}

\begin{document}

\maketitle

\begin{abstract}
In supervised machine learning, the assumption that training data is labelled
correctly is not always satisfied. In this paper, we investigate an instance of
labelling error for classification tasks in which the dataset is corrupted with
out-of-distribution (OOD) instances: data that does not belong to any of the
target classes, but is labelled as such. We show that detecting and relabelling
certain OOD instances, rather than discarding them, can have a positive effect
on learning. The proposed method uses an auxiliary classifier, trained on data
that is known to be in-distribution, for detection and relabelling. The amount
of data required for this is shown to be small. Experiments are carried out on
the FSDnoisy18k audio dataset, where OOD instances are very prevalent. The
proposed method is shown to improve the performance of convolutional neural
networks by a significant margin. Comparisons with other noise-robust
techniques are similarly encouraging.
\end{abstract}


\section{Introduction}
\label{s:intro}

Supervised learning refers to the use of labelled training data, the
availability of which provides a tremendous advantage in many applications of
machine learning. In practice, labels are not always correct
\cite{noise_frenay}, prompting additional efforts to carefully verify them.
This can be prohibitively costly when scaling to large datasets, which often
results in limited data for training. In order to utilise much larger datasets,
there has been interest in learning methods that do not rely on clean data
\cite{fsdnoisy18k_fonseca, noise_sukhbaatar, noise_reed, noise_li}. To this
end, this paper investigates a case of labelling error for audio classification
in which the dataset is corrupted with out-of-distribution (OOD) instances:
data that does not belong to any of the target classes, but is labelled as
such.

Large amounts of annotated data are available to use when considering the world
wide web \cite{freesound_fonseca, audioset_gemmeke}. However, due to the
uncontrolled/miscellaneous nature of these sources of data, irrelevant (OOD)
instances are likely to be encountered when curating the data. For example,
Freesound Annotator \cite{freesound_fonseca} is a platform of datasets
comprised of over \SI{260}{\K} audio samples annotated by the public, where the
authors of this platform have observed a considerable number of OOD instances
\cite{fsdnoisy18k_fonseca}. OOD corruption can occur for a number of reasons,
such as uncertainty in the sound (e.g. being unable to discriminate between
clarinet sounds and flute sounds) and uncertainty in the label semantics (e.g.
`keyboard' could refer to keyboard instruments or it could refer to computer
keyboards).

In this paper, it is argued that certain OOD instances, when labelled
appropriately, can be beneficial for learning, and that this depends on their
likeness to the in-distribution (ID) data. Using a continuous label space, one
can even assign `soft' labels to these instances to reflect uncertainty in what
the most appropriate target class is. By considering the new labels as the
correct labels, OOD corruption can be framed in terms of label noise
\cite{noise_frenay}; for each instance, a (pseudo-)correct label exists, but
the label assigned by the annotator may be incorrect.

Considering the problem in terms of label noise allows the incorporation of
methods developed for label noise. In particular, this paper proposes a
pseudo-labelling method for the OOD training data. There are two main stages:
(1) OOD detection and (2) relabelling. To detect and relabel the relevant
instances, an auxiliary classifier trained on a much smaller dataset of
manually-verified ID examples is used. The original ground truth of the
training data is also exploited. Requiring a small amount of verified data is
not unreasonable, as the cost of doing so is relatively low. Convolutional
neural networks are used as baselines to assess the proposed method and compare
it to alternatives methods. Experiments are carried out on the FSDnoisy18k
dataset \cite{fsdnoisy18k_fonseca}, which is a large audio dataset with a
substantial number of OOD training examples.

\subsection{Related Work}
\label{s:i:related}

The OOD detection methods most relevant to our work are those that use a
discriminative neural network model to directly detect OOD data. Hendryks and
Gimpel \cite{ood_hendryks} proposed using the maximum of the softmax output as
a measure of confidence, where a low value indicates that an instance is OOD.
Using this idea as a baseline, several improvements have been proposed,
including different confidence measures \cite{ood_liang} and changes to the
learning dynamics during training \cite{ood_devries}. Our paper adapts ODIN
\cite{ood_liang}, which applies input pre-processing and logit scaling to
influence the softmax output. There are two key differences, however. First,
the method is modified to detect OOD instances that are \textit{similar} to the
ID data. Second, our paper is concerned with OOD \textit{training} data rather
than \textit{test} data. As a result, we also exploit the ground truth labels.

Pseudo-labelling is a method that has been proposed in the past for the noisy
label problem, with a number of works using neural network models. Reed et al.
\cite{noise_reed} proposed a method called \textit{bootstrapping}, which
adaptively relabels the training data as a combination of the observed label
and the current classifier prediction. Li et al. \cite{noise_li} proposed a
similar method, but instead of using predictions during training, they used the
prediction of an auxiliary classifier trained on a verified dataset. This is
the approach that is adopted in this paper. However, our proposed method also
includes a detection stage to only relabel a subset of the training examples.

\section{Background}
\label{s:background}

A single-label classifier is any function \(f: \Xcal \to \Ycal\) that maps an
instance, \(x \in \Xcal\), to a label, \(y \in \Ycal\), where \(\Ycal \coloneqq
\{1,...,K\}\) and \(K\) is the number of classes. The instance \(x\) is said to
belong to the class corresponding to \(y\). A supervised learning algorithm is
said to train a classifier using \textit{examples} (training data) of the form
\((x, y) \in \Xcal \times \Ycal\) in order to be able to classify future
instances (test data). In a standard learning problem, the training set and
test set are assumed to be sampled from the same distribution \(D\) over
\(\Xcal \times \Ycal\) \cite{noise_van_rooyen}. This is no longer the case when
examples are labelled incorrectly. In the problem that we are studying, the
training set contains instances for which the marginal probability density
function, \(p(x)\), of \(D\) vanishes. Such instances are known as
out-of-distribution instances.

The distribution \(D\) can determine whether an instance is OOD, but it cannot
describe the OOD data itself nor its relation to the ID data. This is limiting,
as OOD instances can possess properties that overlap with the ID data. For
example, clarinets and flutes sound similar regardless of whether one of them
is OOD. In this sense, OOD instances can be `near' or `far' from the ID data.
We argue that this information, which is a manifestation of data uncertainty
\cite{uncertainty_malinin}, can be considered as further knowledge for a
learning algorithm to benefit from. Similar to knowledge distillation
\cite{distillation_hinton}, we use pseudo-labels to convey this knowledge. The
pseudo-labels are generally soft labels such that \(y \in \{z \in
\mathbb{R}_+^K: \norm{z}_1 = 1\}\). This conveys the uncertainty in assigning
an OOD instance to any one class.

From another perspective, assigning soft labels to OOD instances has already
been shown to improve training. Mixup \cite{mixup_zhang} is a data augmentation
technique that linearly combines instances and their labels, such that \(\xh =
\alpha x_1 + (1-\alpha) x_2\) and \(\yh = \alpha y_1 + (1-\alpha) y_2\), where
\(y_1\) and \(y_2\) are understood to be one-hot vectors. When \(y_1 \ne y_2\),
\(\yh\) does not correspond to any one class, meaning that \(\xh\) is OOD
\cite{mixup_guo}. Despite using OOD instances as additional training data,
mixup has been shown to be an effective form of data augmentation
\cite{mixup_zhang, mixup_guo, audio_iqbal}. This is further motivation for our
method. In our problem, however, the pseudo-correct label \(\yh\) is
\textit{not} known and not just the result of a linear combination. To estimate
the pseudo-label, we use two sources of information: the ground truth label and
the prediction of an auxiliary classifier.

\section{Proposed Method}
\label{s:method}

\begin{figure}
  \centering
  \includegraphics[width=0.78\columnwidth]{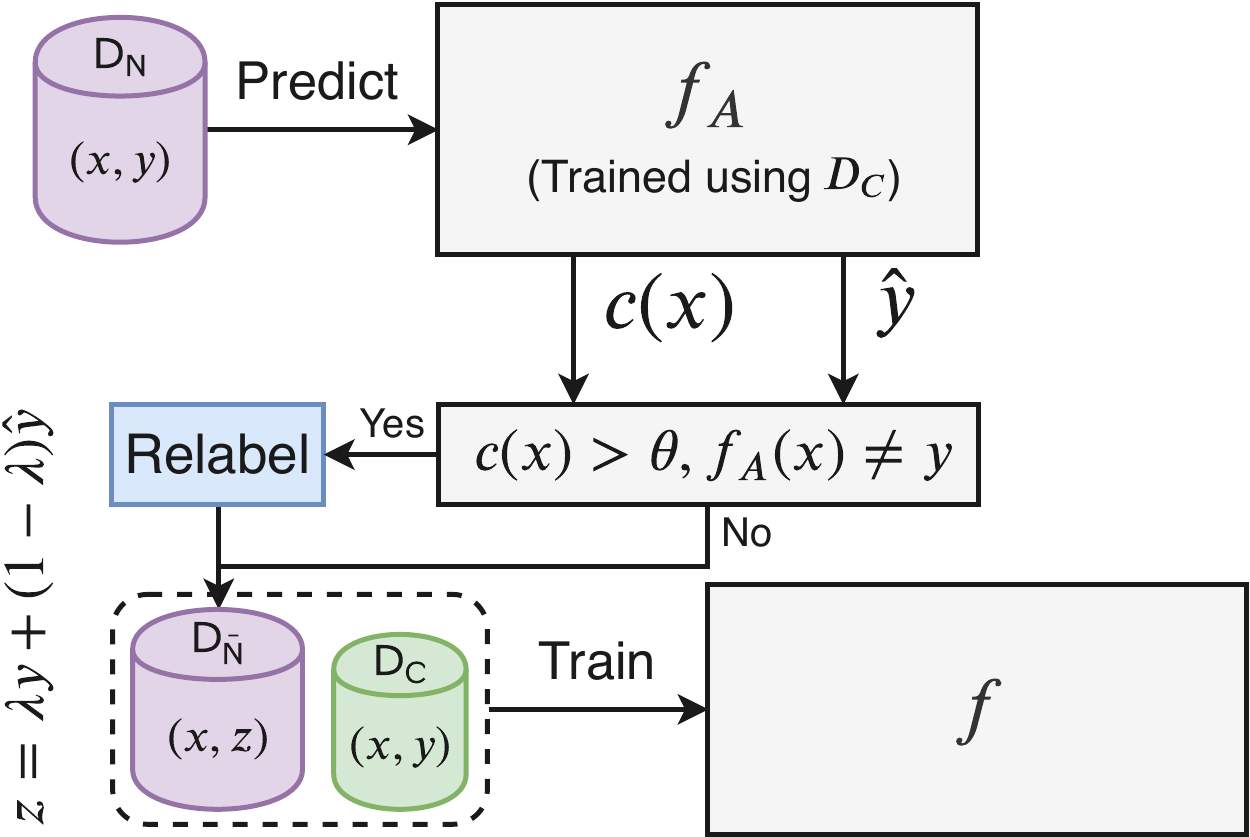}
  \caption{An illustration of how OOD instances are detected and relabelled
    using the auxiliary classifier, \(f_A\), and then used to train the primary
    classifier, \(f\). \(\Dc\), \(\Dn\), and \(\Dr\) denote the datasets
    defined in Section \ref{s:method}.}
  \label{f:decision}
\end{figure}

To detect and relabel OOD instances, another training set, \(\Dc \coloneqq
(\Xsf_C, \Ysf_C)\), consisting of only clean (verified) examples is used. That
is, \(\Dc\) only contains ID instances that are labelled correctly. In
contrast, let \(\Dn \coloneqq (\Xsf_N, \Ysf_N)\) denote the dataset consisting
of noisy (unverified) examples. The general steps of the proposed method are as
follows (cf. Figure \ref{f:decision}):
\begin{enumerate}
  \itemsep -0.15em
  \item Train an auxiliary classifier, \(f_A\), using \(\Dc\).
  \item \label{item:detect} Detect the relevant OOD instances in \(\Dn\) using
    \(f_A\).
  \item \label{item:relabel} Relabel the detected instances using \(f_A\). Let
    \(\Dr\) denote the new dataset after relabelling.
  \item Train the primary classifier, \(f\), using \(\Dc \cup \Dr\).
\end{enumerate}
In the final step, both verified and unverified examples are used to train the
classifier, but with the changes made in step \ref{item:relabel}. The two steps
that deserve the most attention are steps \ref{item:detect} and
\ref{item:relabel}, which are developed in the following subsections.

\subsection{Out-of-Distribution Detection}
\label{s:m:detection}

As outlined in Section \ref{s:i:related}, there has been previous work on OOD
detection based on estimating confidence values for the predictions of the
classifier. Instances for which the confidence, \(c(x) \in [0, 1]\), of the
classifier is lower than a threshold, \(\tau \in (0, 1)\), are considered OOD.
Although these algorithms have been shown to be effective in various test cases
\cite{ood_hendryks}, they have been more successful in detecting instances that
are \textit{far} from the ID data, while OOD instances that are \textit{near}
are typically not detected \cite{ood_liang}. This is not appropriate, as our
relabelling method relies on the OOD instances to be near.

To rectify this, our proposed detection algorithm exploits the availability of
labels in the training data. When there is a \textit{mismatch} between the
label and the classifier prediction, i.e. \(f_A(x) \ne y\) for a training
example \((x, y)\), and the confidence of the classifier is \textit{above} a
threshold, i.e. \(c(x) > \theta\) for \(\theta \in (0, 1)\), the instance \(x\)
is considered as an OOD instance that should be relabelled. Denoting
\(S_{\theta}\) as the set containing these OOD instances, we have
\begin{equation}
  S_{\theta} \coloneqq \{x: c(x) > \theta \text{ and } f_A(x) \ne y\}.
\end{equation}
Selecting instances for which there is a mismatch indicates that the observed
label, \(y\), may be erroneous. By additionally ensuring that \(c(x)\) is high,
it only includes instances that are believed to be similar to one of the target
classes.

One could also incorporate the mismatch condition for low-confidence OOD
instances by defining
\begin{equation}
  S_{\tau} \coloneqq \{x: c(x) < \tau \text{ and } f_A(x) \ne y\}.
\end{equation}
This is intended to detect far-away OOD instances, as with the OOD detection
algorithms in previous work \cite{ood_hendryks, ood_liang, ood_devries}, but
the additional mismatch condition can reduce the number of ID instances that
are incorrectly detected, since a mismatch is less likely for ID instances. The
instances in \(S_{\tau}\) can then be dealt with separately, e.g. by discarding
them. This is explored as an additional step in the experiments presented
later, but is not the main focus of this paper.

\subsection{Pseudo-labelling}
\label{s:m:pseudo}

As explained in Section \ref{s:background}, OOD instances do not belong to any
of the target classes, but they may share properties with one or more of them.
As such, we propose to relabel the instances in \(S_{\theta}\) as a convex
combination of the observed label, \(y\), and the prediction, \(\yh\), given by
the auxiliary classifier.
\begin{equation}
  \label{eq:relabel}
  z = \lambda y + (1-\lambda) \yh.
\end{equation}
In \eqref{eq:relabel}, \(y\) is understood to be a one-hot vector, while
\(\yh\) is a vector of probabilities, i.e. \(\yh \in \{z \in \mathbb{R}_+^K:
\norm{z}_1 = 1\}\), so that \(f_A(x) = \argmax \yh_i\). The parameter \(\lambda
\in (0, 1)\) determines the weight to apply to the observed label and the
prediction. \(\yh\) is used because the auxiliary classifier is confident in
the prediction, so it is likely to be an optimal label for learning. On the
other hand, there is a chance that it is wrong or suboptimal relative to the
observed label. The weight, \(\lambda\), can be interpreted as the prior belief
that this is the case.

This method of relabelling has also been proposed in the past for noisy labels
\cite{noise_li}, where the authors found that it was an effective approach for
several noisy datasets. For the selection of the parameter \(\lambda\), they
used a heuristic that derived the weight as a function of the performance of
the auxiliary classifier on clean data and noisy data.

\section{Experiments}
\label{s:experiments}
  
In this section, experimental results are presented to evaluate the proposed
method\footnote{Source code: \url{https://github.com/tqbl/ood_audio}}. The
experiments were carried out using the FSDnoisy18k dataset
\cite{fsdnoisy18k_fonseca}, which is a crowdsourced audio classification
dataset created as part of Freesound Annotator \cite{freesound_fonseca}. It
contains \num{18532} audio clips across \num{20} classes, totalling \num{42.5}
hours of audio. The clip durations range from \SI{300}{\ms} to \SI{30}{\s}. The
dataset is divided into a training set containing \num{17585} clips and a test
set containing \num{947} clips. The test set has been manually verified so that
all of the audio clips are ID. In contrast, only \SI{10}{\%} of the training
set is verified: this is the data that is used to train the auxiliary
classifier. It has been estimated \cite{fsdnoisy18k_fonseca} that \SI{45}{\%}
of the unverified labels are incorrect, and that \SI{84}{\%} of the incorrect
labels are OOD.

\subsection{Baseline System}
\label{s:r:baseline}
The evaluated systems are based on two baseline convolutional neural networks
(CNNs) \cite{vgg_simonyan, densenet_huang}. We used two baselines to
investigate two different settings: using a randomly-initialised model and
using a pre-trained model. The two baselines are \textit{VGG9} (randomly
initialised) and \textit{DenseNet-201} (pre-trained). VGG9 is based on VGG
\cite{vgg_simonyan, audio_iqbal} and contains \num{8} convolutional layers and
\num{1} fully-connected layer. DenseNet-201 is DenseNet with 201 layers
\cite{densenet_huang} and was pre-trained using ImageNet \cite{imagenet_deng}.
Although ImageNet is an image dataset, we found that it was surprisingly
effective for pre-training. The architecture of each baseline was chosen
independently based on performance. DenseNet-201 performed better than VGG9
when pre-training with ImageNet, while VGG9 performed better than DenseNet-201
when using randomly-initialised weights.

Features were extracted by converting the audio waveforms into logarithmic
mel-spectrograms (log-mels) with a window length of \num{1024}, a hop length of
\num{512}, and \num{64} bands. To ensure the neural networks received
fixed-length inputs, we used a block-based approach as used in our previous
work \cite{audio_iqbal}. That is, the feature vectors were partitioned into
blocks of length \num{128} and processed independently.

Models were trained using the categorical cross-entropy loss function with the
Adam optimization algorithm \cite{adam_kingma}. Training was carried out for
\num{40} epochs with a batch size of \num{128} and a learning rate of
\num{0.0005}, which was decayed by \SI{10}{\%} after every two epochs. The
primary classifier, \(f\), and the auxiliary classifier, \(f_A\), were trained
identically. However, \(f_A\) was specifically trained using the pre-trained
DenseNet-201 model, regardless of the model used for \(f\).

Unless otherwise stated, all hyperparameter values were selected by evaluating
the models with a validation set, which contained \num{15} manually-verified
examples from each class, and was sampled from the training set.

\subsection{Evaluated Systems}
\label{s:e:systems}
Several systems were evaluated to assess the performance of the proposed
method. Each system applies to both baselines. We evaluated a number of
variations of the proposed method (starred below) as well as alternative
methods proposed for noise robustness in general. The systems are as follows:
\begin{itemize}
  \itemsep -0.25em
  \item \textit{Clean}: The baseline trained with clean examples only.
  \item \textit{Clean-DA}: \textit{Clean} with data augmentation. The
    DenseNet variant of this system was used to train \(f_A\).
  \item \textit{Baseline}: The baseline trained with all of the examples.
  \item \textit{\supstar{OOD-R}}: The method proposed in this paper.
  \item \textit{\supstar{OOD-RD}}: Equivalent to \textit{OOD-R},
    except instances in \(S_{\tau}\) (c.f. Section \ref{s:m:detection}) are
    discarded.
  \item \textit{\supstar{All-R}}: All the examples in \(\Dn\) are
    relabelled.
  \item \textit{Bootstrap}: Labels are updated dynamically using the
    bootstrapping method \cite{noise_reed}. No OOD detection.
  \item \textit{\(\Lq\) Loss}: The baseline system with the \(\Lq\) loss
    \cite{lq_zhang} (with \(q=0.7\)) instead of the cross-entropy loss.
  \item \textit{Noise Layer}: An additional linear layer maps
    the predictions to the same noisy space as the observed labels
    \cite{noise_sukhbaatar}. This layer is removed during inference.
\end{itemize}

The proposed system and its variations were configured with \(\tau=0.4\),
\(\theta=0.55\), and \(\lambda=0.5\). The value of \(\lambda\) was selected
based on the heuristic given in \cite{noise_li}. The values of \(\theta\) and
\(\tau\) were selected by applying the detection algorithm on the validation
set and selecting the threshold for which less than \SI{5}{\%} of the instances
were (incorrectly) detected. This way, the selection of the thresholds is
interpretable and independent of \(f_A\). Instead of using the softmax output
directly, ODIN \cite{ood_liang} was used to compute \(c(x)\); we found that
ODIN detected more instances in \(\Dn\) for a given number of incorrectly
detected validation instances. Using ODIN, \SI{25.7}{\%} (resp. \SI{13.8}{\%})
of \(\Dn\) was detected as belonging to \(S_{\theta}\) (resp. \(S_{\tau}\)).

The purpose of \textit{Clean-DA} is to compare data augmentation to using the
noisy examples. We experimented with mixup \cite{mixup_zhang} and SpecAugment
(time/frequency masking) \cite{specaug_park}, and adopted the latter as it gave
superior performance. The \(\Lq\) loss is designed to be robust against
incorrectly-labelled data, and is the approach taken by the authors of
FSDnoisy18k \cite{fsdnoisy18k_fonseca}. The bootstrapping method is similar to
the proposed method, except there is no OOD detection and no auxiliary
classifier, as examples are relabelled \textit{during} training as a
combination of the ground truth and \(f\)'s current prediction. The noise layer
was proposed for class-conditional label noise \cite{noise_sukhbaatar}, and was
utilised by Singh et al. \cite{fsdnoisy18k_singh} for FSDnoisy18k.

To score the performance of the systems, average precision (AP) and accuracy
were used as metrics. Both were computed as micro-averages and reported in
percentages, where a higher percentage means higher performance. Five trials
were carried out for each experiment to account for uncertainty.

\subsection{Results}
\label{s:r:results}

\begin{table}[t]
  \setlength{\tabcolsep}{2pt}
  \renewcommand{\arraystretch}{0.96}

  \caption{Experimental results for all systems. AP and accuracy are reported
    with \SI{68}{\%} confidence intervals.}
  \label{t:main_results}
  \centering
  \begin{tabularx}{0.48\textwidth}{lXXXX}
    \toprule
                     & \multicolumn{2}{c}{AP}                    & \multicolumn{2}{c}{Accuracy} \\
    System           & VGG                 & DenseNet            & VGG                 & DenseNet \\
    \midrule
    Clean            & \score{71.8}{0.21}  & \score{72.3}{0.47}  & \score{67.5}{0.23}  & \score{67.5}{0.23} \\
    Clean-DA         & \score{74.6}{0.39}  & \score{75.7}{0.34}  & \score{66.4}{0.45}  & \score{69.4}{0.24} \\
    \midrule
    Baseline         & \score{81.6}{0.55}  & \score{84.8}{0.25}  & \score{74.5}{0.71}  & \score{78.0}{0.36} \\
    \supstar{OOD-R}  & \scoreb{86.0}{0.15} & \scoreb{88.0}{0.21} & \scoreb{77.9}{0.14} & \scoreb{81.0}{0.50} \\
    \supstar{OOD-RD} & \scoreb{86.1}{0.37} & \score{87.1}{0.36}  & \scoreb{78.0}{0.56} & \score{80.3}{0.15} \\
    \supstar{All-R}  & \score{84.5}{0.55}  & \scoreb{88.1}{0.19} & \score{77.1}{0.68}  & \scoreb{81.4}{0.27} \\
    Bootstrap        & \score{81.2}{0.36}  & \score{84.9}{0.60}  & \score{74.8}{0.80}  & \score{78.7}{0.46} \\
    \(\Lq\) Loss     & \score{83.3}{0.48}  & \score{85.0}{0.72}  & \score{76.4}{0.56}  & \score{78.6}{0.64} \\
    Noise Layer      & \score{83.0}{0.47}  & \score{84.7}{0.82}  & \score{76.2}{0.34}  & \score{78.0}{0.75} \\
    \bottomrule
  \end{tabularx}
\end{table}

The results are presented in Table \ref{t:main_results}. When training with the
clean examples only, data augmentation resulted in a noticeable improvement in
performance. However, this improvement can be seen to be relatively small
compared to using all of the examples for training. This shows that there is a
benefit to training with the noisy examples in this dataset, despite a large
percentage of them being OOD/incorrect. Among the alternative methods, the
\(\Lq\) loss and noise layer both resulted in large improvements, but only for
the VGG baseline. Bootstrapping did not improve the performance for either
baseline.

The proposed method, \textit{OOD-R}, performed the best, with significant gains
seen for both baselines. \textit{OOD-RD}, which also discards examples in
\(S_{\tau}\), did not perform better than \textit{OOD-R}; discarding the
examples actually worsened the performance for the DenseNet model, possibly due
to removing ID examples. These results suggest that the neural networks are
robust to far-away OOD instances being present in the training set, and that
removing them should not be a priority. \textit{All-R}, which relabels all of
the examples in \(\Dn\), did not perform as well as \textit{OOD-R} for the VGG
model, which demonstrates the importance of the detection stage. On the other
hand, there was no discernible difference between \textit{All-R} and
\textit{OOD-R} for the DenseNet model. A possible reason is that the
pre-trained DenseNet model is more robust to the label noise introduced when
relabelling the low-confidence examples.

\section{Conclusion}
\label{s:conclusion}

In this work, we investigated the problem of learning in the presence of
out-of-distribution (OOD) data. We argued that OOD data can possess properties
that are characteristic of the target classes, so that appropriately
relabelling the instances to reflect this, rather than discarding them, can
benefit training. Our proposed method involved training an auxiliary classifier
on a smaller verified dataset, and using its predictions, along with the ground
truth labels, to detect and relabel the relevant OOD instances. Using
convolutional neural network baselines, experiments with the FSDnoisy18k
dataset showed that our method substantially improves performance. The results
also suggested that OOD instances that are very different from the target
classes have little effect on performance when present in the training data.
Future work includes investigating other detection and pseudo-labelling
methods, including those that do not require any verified data.

\bibliographystyle{IEEEbib}
\bibliography{references}

\end{document}